\documentclass[fleqn,usenatbib]{mnras}

\usepackage{newtxtext,newtxmath}

\usepackage[T1]{fontenc}
\usepackage{lscape}
\usepackage{comment}
\DeclareRobustCommand{\VAN}[3]{#2}
\let\VANthebibliography\thebibliography
\def\thebibliography{\DeclareRobustCommand{\VAN}[3]{##3}\VANthebibliography}

\usepackage{graphicx}	
\usepackage{amsmath}	

\usepackage[dvipsnames]{xcolor}
\usepackage[colorinlistoftodos]{todonotes}
\definecolor{salmon}{rgb}{0.95,0.5,0.25}

\title[Off-centre black holes in BCGs]{Off-centre supermassive black holes in bright central galaxies}

\author[Chu et al.]{
Aline Chu $^{1}$\thanks{E-mail: chu@iap.fr},
Pierre Boldrini $^{1,2}$\thanks{E-mail: boldrini@iap.fr} and Joe Silk $^{1,3,4}$
\\
$^{1}$Sorbonne Universit\'e, CNRS, UMR 7095, Institut d'Astrophysique de Paris, 98 bis bd Arago, 75014 Paris, France \\
$^{2}$Universit\'e de Lorraine, CNRS, Inria, LORIA, F-54000 Nancy, France\\
$^{3}$Department of Physics and Astronomy, The Johns Hopkins University, Baltimore MD 21218, USA\\
$^{4}$Beecroft Institute for Particle Astrophysics and Cosmology, Department of Physics, University of Oxford, Oxford OX1 3RH, UK\\}

\date{Accepted XXX. Received YYY; in original form ZZZ}

\pubyear{2015}

\begin{document}
\label{firstpage}
\pagerange{\pageref{firstpage}--\pageref{lastpage}}
\maketitle

\begin{abstract}

Supermassive black holes (SMBHs) are believed to reside at the centre of massive galaxies such as brightest cluster galaxies (BCGs). However, as BCGs experienced numerous galaxy mergers throughout their history, the central BH can be significantly kicked from the central region by these dynamical encounters. By combining the TNG300 simulations and orbital integrations, we demonstrate that mergers with satellite galaxies on radial orbits are a main driver for such BH displacements in BCGs. BHs can get ejected to distances varying between a few parsecs to hundreds of kiloparsecs. Our results clearly establish that SMBH offsets are common in BCGs and more precisely a third of our BHs are off-centred at $z=0$. This orbital offset can be sustained for up to at least 6 Gyr between $z=2$ and $z=0$ in half of our BCGs. Since the dense gas reservoirs are located in the central region of galaxies, we argue that the consequences of off-center SMBHs in BCGs are to quench any BH growth and BH feedback.

\end{abstract}




\section{Introduction}

Galaxy clusters are the largest virialized structures known today. These rich systems of hundreds to thousands of galaxies are the perfect targets to study how environment can impact the formation of galaxies and their evolution. Located at the intersection of the filaments which compose the cosmic web, clusters are believed to form mainly by accreting small galaxies and groups which fall along the filaments, or by mergers with other systems \citep[][]{Kravtsov2012}. Galaxies, gas clumps and globular clusters, which are captured by the cluster potential, would sink to the central region of the cluster. Indeed, these satellites are bound to merge with the central galaxy, which is often referred to as the Brightest Cluster Galaxy (BCG) \citep{1999MNRAS.306..857C,2007AJ....133.1741B}. 

BCGs are among the most massive galaxies observed in the Universe \citep[][]{Dubinski_1998}. These peculiar elliptical galaxies, which generally lie at the bottom of the cluster potential well, are more likely to encounter other objects due to their special location, and thus, to undergo many mergers during their lifetimes which promotes their growth. BCGs form via environmental processes \citep[][and references therein]{castignani2020molecular}, and suffer and reflect all the processes which formed and shaped their host clusters. Hence, BCGs have properties which are closely linked to those of their host clusters \citep[see][]{lauer2014brightest,2022ApJ...938....3S}.

In the local Universe, black holes (BHs) are nearly ubiquitous in galaxies \citep{2013ARA&A..51..511K,2016ApJ...818...47S,2016ApJ...831..134V}. In particular, BCGs host among the most massive BHs ever detected, the supermassive black holes (SMBH). Recently, measurements were reported of SMBH masses in excess of $10^{10}$ M$_{\sun}$ \citep{2011Natur.480..215M,2012MNRAS.424..224H,2015ApJ...805...35H,2013ApJ...764..184M,2016Natur.532..340T}. We expect the central BHs to grow at the same time as their host galaxies by accreting matter brought in by infalling galaxies. That is the reason why central BH masses seem to be correlated with the stellar and bulge masses of their host galaxies \citep{1998AJ....115.2285M,2009ApJ...698..198G,2013MNRAS.435.3085M, McConnell_2013,Sahu_2019}. A prominent example is the SMBH of Holm 15A, the BCG of the galaxy cluster Abell 85, which is the most massive BH directly detected via stellar dynamics with a mass of about $4\times10^{10}$ M$_{\sun}$, i.e. 2$\%$ of the total stellar mass of the central galaxy \citep{2019ApJ...887..195M}.

However, it has been demonstrated that BHs are not necessarily located exactly at the bottom of the cluster potential, inside and at the center of the central galaxy. In particular, tens of thousands of active galactic nuclei (AGN) are not located at the centers of their host galaxies. These observations exhibit BH offsets between tens of parsecs to a few kiloparsecs \citep{2014ApJ...796L..13M,2016ApJ...817..150M,2020ApJ...888...36R,2019ApJ...885L...4S}. Many scenarios have been proposed to explain these off-centre BHs \citep[][]{2010PhRvD..81j4009S, 2018ApJ...857L..22T, 2009ApJ...690.1031B,2014ApJ...789..112C, 2005LRR.....8....8M,2005MNRAS.358..913V,2007PhRvL..99d1103L,2012AdAst2012E..14K,boldrini2020}. It is important to notice that most of these scenarios need to invoke interactions and mergers with other galaxies or dark matter (DM) substructures \citep{2019MNRAS.482.2913B,2019MNRAS.486..101P,bellovary2021_10.1093/mnras/stab1665,bellovary2019_10.1093/mnras/sty2842,boldrini2020}. 

This BH feature was also found in more massive galaxies such as Andromeda (M31). It has been shown that its SMBH is offset by 0.26 pc from the centre of the galaxy \citep{1999ApJ...522..772K}. As there are several indications of a recent merger activity in M31, it was demonstrated that the accretion of a satellite on a highly eccentric orbit naturally explains this off-centre BH in our neighbor galaxy \citep{2020MNRAS.498L..31B}. As BCGs are the results of numerous galaxy mergers throughout their history, we suspect that some mergers could have significantly offset the SMBH located at the center of BCGs to distances even larger than of the order of a kiloparsec.

In this paper, we aim to examine the impact of mergers between BCGs and satellite galaxies, within galaxy clusters, on the SMBH of BCGs. In fact, we track the orbit of the central SMBH in 370 BCGs by using orbital integration methods after analysing the merger history of the central galaxies provided by the TNG300 simulation. The paper is organized as follows. Our method is based on the coupling between TNG300 data and orbital integration methods and is described in details in Section~\ref{section:method}. We present our results in Section~\ref{section:results}, and discuss the implications of this BH behaviour in Section~\ref{section:discussion}. Section~\ref{section:conclusion} presents our conclusions and perspectives.

\section{METHOD}
\label{section:method}

Using a large sample of 370 galaxy clusters detected by \citet{2018MNRAS.481.1809B} in Illustris-TNG300 \citep[][]{Nelson+19a,Pillepich_2017,Springel_2017,Nelson_2017,Naiman_2018,Marinacci_2018}, we track the orbit of a test SMBH throughout the merger history of the BCG, by integrating its orbit in the fixed potential of the BCG using the publicly available code \texttt{galpy}\footnote{Available at \url{https://github.com/jobovy/galpy}} \citep{2015ApJS..216...29B}. The sample is limited to masses $M_{500}$ $>$ 10$^{\rm 13.75}$ M$_\odot$ and has a median mass $M_{500}$ = 8.8 $\times$ 10$^{\rm 13}$ M$_\odot$ at $z$ = 0. The following describes our approach. 

\subsection{Merger history of BCGs with TNG300}

For each cluster in TNG300, we identify the BCG at $z=0$, which is assumed to be the most massive galaxy in the cluster at $z=0$. Going back through its merger history tree until $z=2$, we search for all the main mergers that the BCG has undergone. Indeed, BCGs were shown to have photometric properties and shapes that do not change much in that redshift range, hinting at an earlier formation epoch \citep{2021A&A...649A..42C,Chu+22}. Moreover, other studies such as \citet{delucia2007} and \citet{thomas10.1111/j.1365-2966.2010.16427.x} have shown evidence that the stellar population in such galaxies has most likely settled since $z=2$ as most of the stars in BCGs were already formed in-situ a long time ago, and the stellar masses measured today are found not to be significantly different from their masses 9 billion years ago \citep{collins2009}. More precisely, BCGs only grow a factor of 1.8 in mass between $z=1$ and $z=0$ \citep{2013MNRAS.434.2856B}.

We identify the BCG at snapshot $n$ and go back one snapshot to $n-1$ to look for its first progenitor (FP) (the most massive galaxy in the cluster at snapshot $n-1$). If a merger happened, we also identify the most massive satellite galaxy with which the BCG has merged, which is referred to as the next progenitor (NP) in the TNG300 simulation. At snapshot $n-1$, the NP will also have the BCG as a descendant. We thus retrieve all FPs and NPs at each snapshot present in the parent halo (or cluster). If two galaxies are merging, we retrieve their positions, velocities, half-mass radii, stellar and DM masses, which will be used to integrate the orbit of the satellite in the potential of the BCG. We find that BCGs in this sample have undergone between 12 and 130 mergers since $z = 2$.

\subsection{Orbital integrations with \texttt{galpy}}\label{dfsection}
\

For each satellite galaxy (or NP), we integrate its orbit forward in time in the potential of the BCG (or FP) from $z = 2$ until $z = 0$. The potential of the BCG consists in a Hernquist and Plummer profile for the DM and stellar components respectively, which is computed considering the properties found in TNG300. All masses and scale radii are taken from the cosmological simulation. The DM and stellar masses of the BCG remain fixed along all the orbital integrations. In fact, we neglect the mass contribution of satellites from galaxy mergers. The satellite is assumed to be a point mass, which suffered the dynamical friction from the BCG. Its total half-mass radius $r^{\rm NP}_{\rm hm}$ is incorporated in the dynamical friction force. We used the Chandrasekhar dynamical friction force, which is already implemented in \texttt{galpy} and roughly follows \cite{2016MNRAS.463..858P}. The impact of the dynamical friction modeling on our findings is discussed in Appendix~\ref{appDF}. Besides, we also neglect the mass loss of the satellite, which could reduce the impact of the merger with the BCG. The assumption that the satellite mass remains constant during integrations is addressed in Appendix~\ref{appML}.

\begin{figure}
\centering
\includegraphics[width=\hsize]{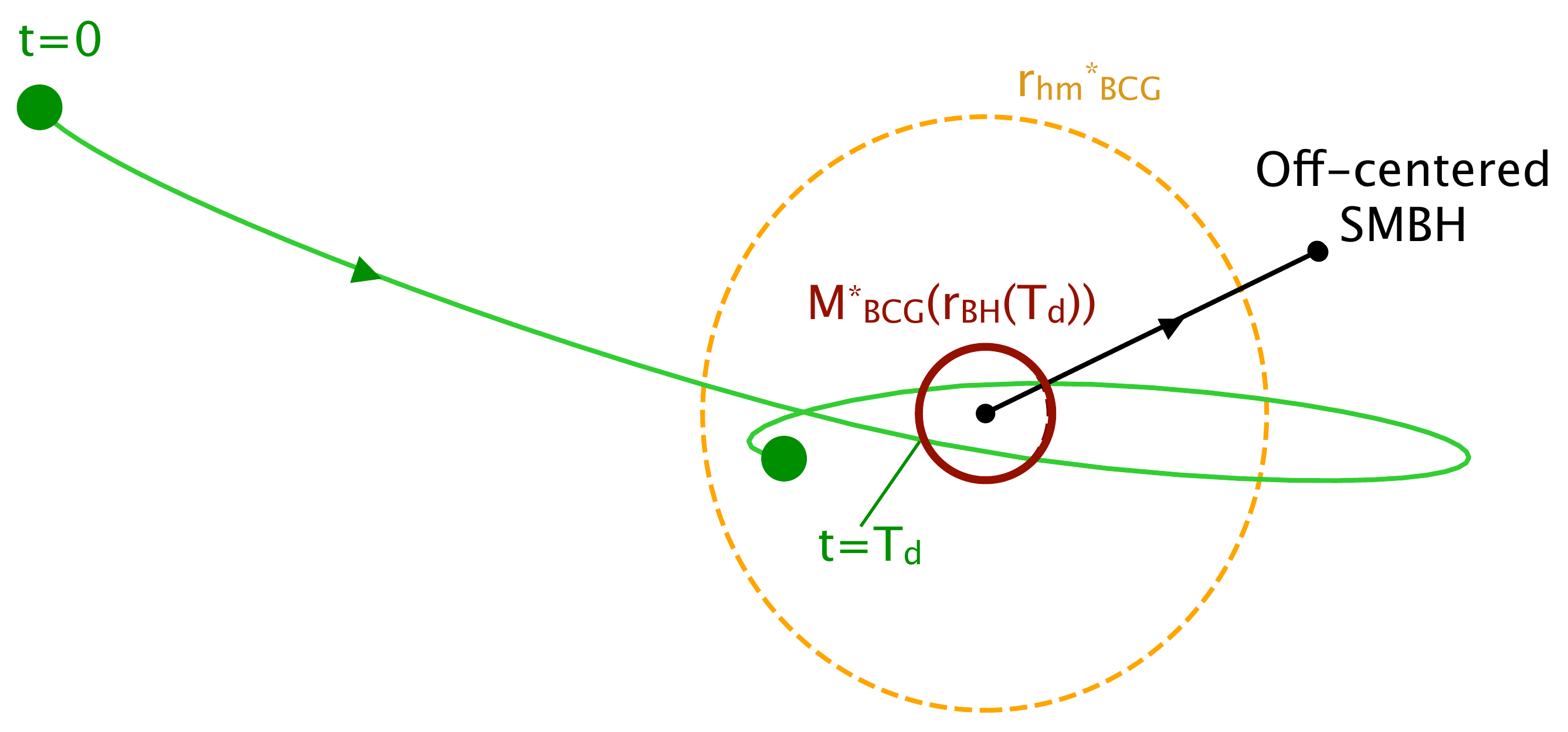}
\caption{\textit{Dynamical heating from satellites of BCG:} Scheme illustrating the radial orbit of a satellite (green points), falling into the potential of the BCG, which kicks away the central BH (black point) after its passage. It results in an off-centered SMBH. The maroon circle represents the region of the BCG where Equations~\eqref{eq1} and ~\eqref{eq2} are satisfied. We note $T_{d}$ the time at which both of these conditions are met. The dotted orange circle represents the half-mass radius of the BCG stellar component.}
\label{fig0}
\end{figure}

\begin{figure}
\centering
\includegraphics[width=\hsize]{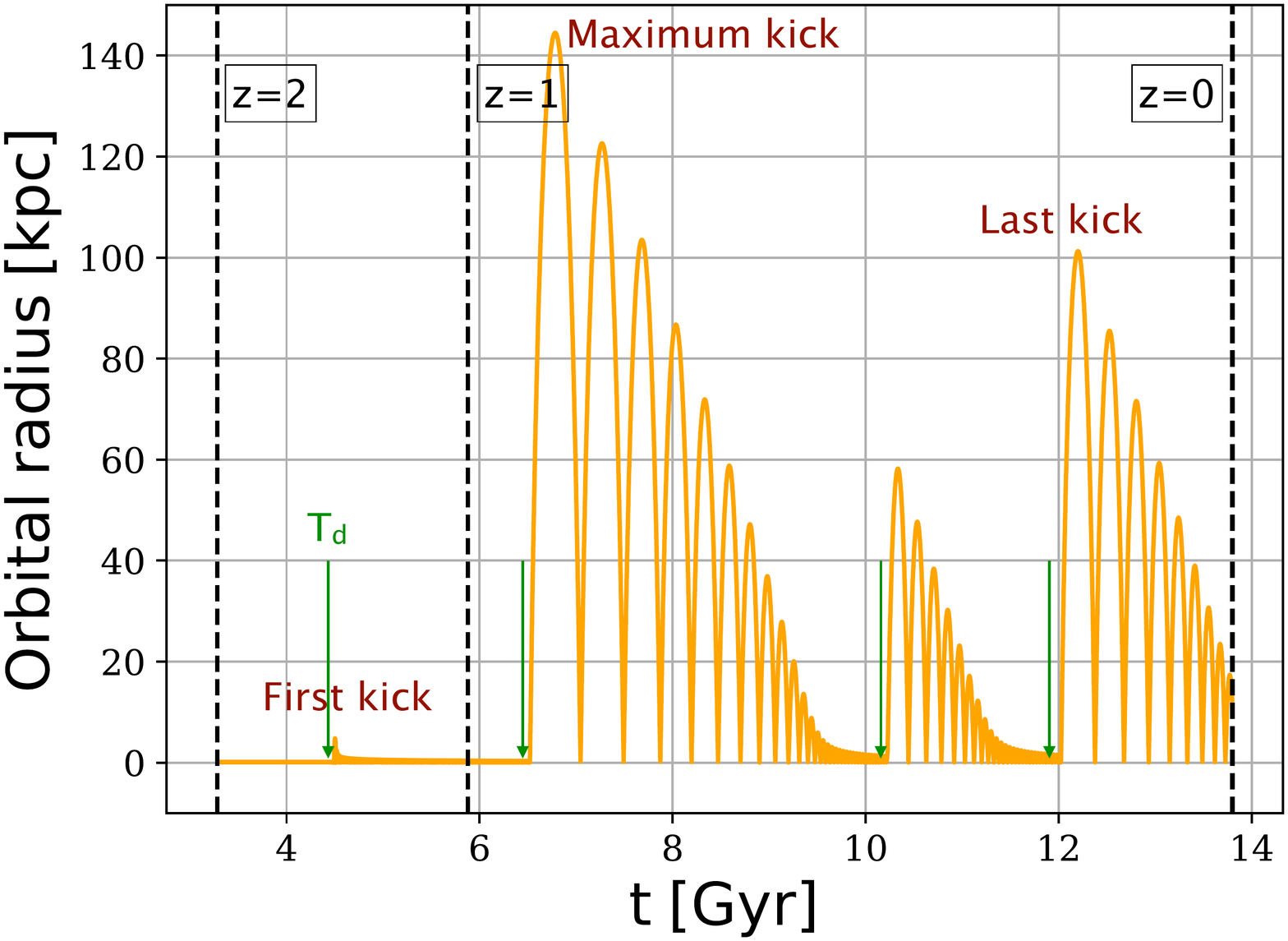}
\caption{\textit{Black hole kicks in BCGs:} Orbital radius of a M$_{\rm SMBH} = 2.6\times10^{10}$ M$_{\sun}$ SMBH in a BCG potential composed of a DM halo of M$_{\rm DM} = $8.1$\times10^{12}$ M$_{\sun}$ and a stellar component of M$_{*} = $4.4$\times10^{11}$ M$_{\sun}$ between $z = 2$ and $z = 0$. The SMBH has experienced 17 mergers since $z=2$ but only 4 mergers have satisfied our kicking criteria described by Equations~\eqref{eq1},~\eqref{eq2} and ~\eqref{eq5}, showed by green arrows. The BH can be ejected to hundreds of kiloparsecs. Between mergers, we observe the orbital decay of the BH due to dynamical friction. At $z = 0$, the BH is still significantly off-centered by about 10 kpc. }
\label{fig1}
\end{figure}

We aim to determine if the satellite galaxy, via the merger, dynamically heats the central region of the BCG and more particularly its SMBH. The satellite needs to be massive enough and pass close enough to the center of the BCG on its first orbit to be able to transfer energy to the central region of the galaxy during the merger. Two conditions need to be met:

\begin{itemize}

        \item The distance $d$ between the centers of the satellite and the BCG must be smaller than the half-mass radius of the NP $r^{\rm sat}_{\rm hm}$: 
        \begin{equation}
           d \leq r^{\rm sat}_{\rm hm}.
            \label{eq1}
        \end{equation}
        This condition needs to be true on the satellite's first passage in the potential of the BCG (radial merger). Indeed, through each orbit, the satellite will transfer its mass to the total mass of the BCG with which it is merging until all its mass has been consumed by the BCG. In order to simplify the problem, we neglect the mass loss of the satellite, which is valid only on its first passage.
        
        \item The total mass of the satellite galaxy must be bigger than the total mass of the FP contained in a radius $d$:
        \begin{equation}
            M^{\rm sat}_{\rm tot} \geq M^{\rm BCG}_{\rm int}(d). 
            \label{eq2}
        \end{equation}
        Only under this condition, the BCG potential can be displaced due to the mass deposit from the satellite \citep{2010ApJ...725.1707G,2006MNRAS.366..429R}. Then the BH can vacate the central region of the galaxy. In our mass calculations, we take into account the DM and stellar components. 
        
\end{itemize}

The two previous conditions are equivalent to checking if the satellite has a radial orbit. Indeed, the satellite must pass by the center or very close to the center of the BCG in order to affect the central SMBH. We note $T_{d}$ the time at which both of these conditions are met. We update at each $T_{d}$ the stellar and DM masses of the BCG given by the TNG300 simulation.
Figure~\ref{fig0} illustrates a satellite which falls in the potential of the BCG passing by its center, in a radius $r_{\rm hm}^{\rm sat}$ as described in Equation~\ref{eq1}. The SMBH is significantly kicked by the satellite galaxy (see Figure~\ref{fig0}).

Using the simulation information, we are able to follow the orbit of satellites throughout the merger history of the BCG with \texttt{galpy}, by considering their initial positions, velocities, and masses. Then, a SMBH with a constant mass is placed at the centre of each BCG potential. As we are not using the BH as a PartType5 particle from the simulation, the mass of the SMBH was calculated based on the \cite{1998AJ....115.2285M} relation: M$_{\rm SMBH}$ = 0.06 M$^{\rm BCG}_{*}$, with M$^{\rm BCG}_{*}$ the stellar mass of the BCG at the first T$_{\rm d}$. We suppose that the SMBH is initially on a circular orbit at a arbitrary radius, which is 100 times its Schwarzschild radius, as the SMBH should lie in the central region of its host galaxy. After a careful check, choosing a smaller radius does not significantly affect our results. If both Equations ~\eqref{eq1} and ~\eqref{eq2} are satisfied, the SMBH is kicked at T$_{\rm d}$ with a velocity based on the analytic expansion derived in \cite{2009ApJ...699L.178N}:
\begin{equation}
    v^{\rm first}_{\rm kick} = \sqrt{\frac{(1 + \eta \epsilon ^2)}{(1 + \eta)}} v^{\rm SMBH}_{\rm c}(100\times R_{\rm s}),
    \label{eq3}
\end{equation}
with $v^{\rm SMBH}_{\rm c}$, which is the circular velocity of the SMBH at 100 times its Schwarzschild radius $R_{\rm s}$,
\begin{equation}
 \eta = M^{\rm sat}_{\rm tot}/M^{\rm BCG}_{\rm int}(d), 
 \label{eq4}
\end{equation}
and
\begin{equation}
\epsilon = \frac{v^{\rm sat}}{v^{\rm SMBH}_{\rm c}},
\end{equation}
where $v^{\rm sat}$ the velocity norm of the satellite galaxy at T$_{\rm d}$ (see Figure~\ref{fig1}). Our criterion~\eqref{eq2} is established to maximise the energy transferred to the SMBH via the velocity kick, described by Equations~\eqref{eq3} and ~\eqref{eq4}. All the velocities are calculated relative to the BCG. We then integrate the BH orbit with its new velocity by applying dynamical friction to it with a constant mass.

At each subsequent merger, we check if the SMBH, which might have been kicked from the center of the galaxy, passes close to the inner region of the BCG heated by the satellite. We, therefore, use the orbital radii of the SMBH and distance $d$ between the centers of the satellite and the BCG to define a simple criterion in order to establish if the BH can be affected after this new merger: 
\begin{equation}
r^{\rm SMBH}_{\rm orb} \leq d. 
 \label{eq5}
\end{equation}
In order to satisfy this condition, the BH must have had sufficient time to come back to the BCG centre before the next merger happens. Then, the SMBH is kicked once again with a new velocity: 
\begin{equation}
    v^{\rm sub}_{\rm kick} = \sqrt{\frac{(1 + \eta \epsilon)}{(1 + \eta)}}v^{\rm SMBH}_{\rm N}. 
\end{equation}
The only difference with Equation~\eqref{eq3} is that the SMBH is no longer in a circular orbit and now has a velocity norm $v^{\rm SMBH}_{\rm N}$. 

First, we plot the orbit of the SMBH within a BCG which went through multiple mergers throughout its history. As an example, Figure~\ref{fig1} depicts the four kicks that a M$_{\rm SMBH}=2.6\times10^{10}$ M$_{\sun}$ BH has experienced in a BCG potential composed of a DM halo of M$_{\rm DM}=$8.1$\times10^{12}$ M$_{\sun}$ and a stellar component of M$_{*}=$4.4$\times10^{11}$ M$_{\sun}$ between $z = 2$ and $z = 0$. The maximum distance reached by the SMBH is about 150 kpc. The SMBH has been heated by a first merger between $z=2$ and $z=1$ and by three more mergers between $z=1$ and $z=0$. We emphasize that the number of mergers shown in Figure~\ref{fig1} may not be equal to the total number of mergers experienced by the BCG. Indeed, the BCG has experienced 17 mergers since $z=2$ but only 4 mergers have satisfied our kicking criteria described by Equations~\eqref{eq1},~\eqref{eq2} and ~\eqref{eq5}. In this example, the BCG undergoes its last merger almost 2 Gyr ago, the SMBH was kicked up to 100 kpc away from the center, which is to say beyond $r^{\rm BCG}_{\rm hm}=32$ kpc at $z=0$. After being kicked, we observe the orbital decay of the SMBH in the BCG potential due to dynamical friction. We argue that the SMBH needs to return to the central region to be kicked again (see Figure~\ref{fig1}). Chances of encounter between an offcenter BH and a falling satellite are very small. In the case of dynamical interactions out of the central regions of BCGs, the collision would have little impact. Indeed, it is really hard to satisfy the criterion ~\eqref{eq2} in these regions of BGCs as the major impact of satellites takes place especially in the central region where the enclosed mass of BCG is smaller (at large distances d, M$^{\rm sat}_{\rm tot}$ would need to be comparable to the total mass of the BCG M$^{\rm BCG}_{\rm int}(d)$). We stress that at $z = 0$, the BH is still significantly off-centered by about 10 kpc. 

We apply this procedure to our full sample of 370 galaxy clusters from TNG300.

\section{Results}
\label{section:results}

\begin{figure}
\centering
\includegraphics[width=\hsize]{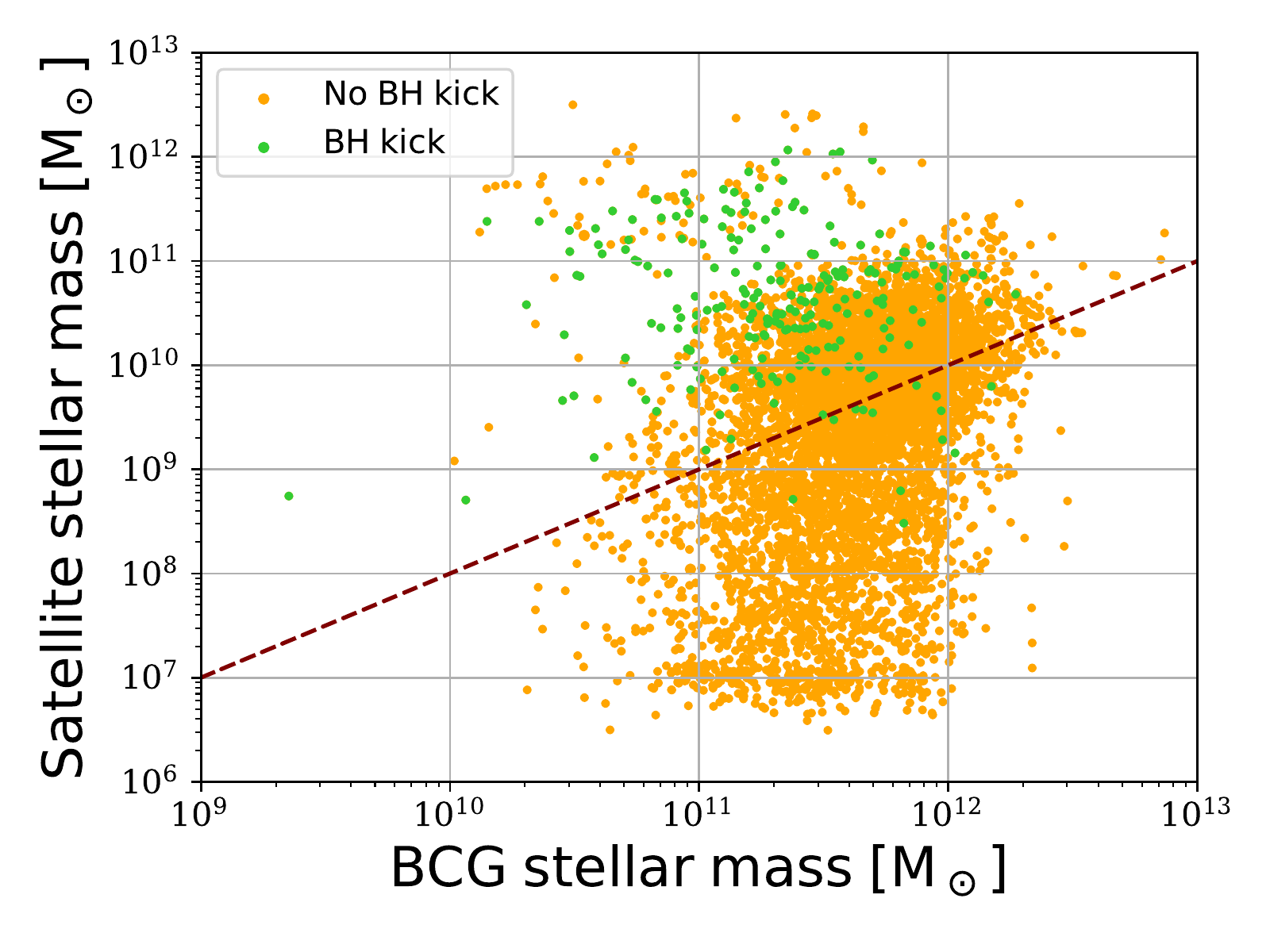}
\caption{\textit{Mergers in our full sample:} Satellite stellar mass as a function of the BCG stellar mass. We distinguish two different categories of mergers in our full sample of 370 BCGs: mergers which have kicked the central BH (green points, 229 mergers), and mergers which have not affected the BH (orange points, 6628 mergers). The dotted line represents a constant ratio of 100 in mass. The majority of mergers leave the BH unaffected. BHs are mainly kicked by satellites which have stellar masses $M^{\rm sat}_{*}>M^{\rm BCG}_{*}/100$.}
\label{fig1a}
\end{figure}

\begin{figure}
\centering
\includegraphics[width=\hsize]{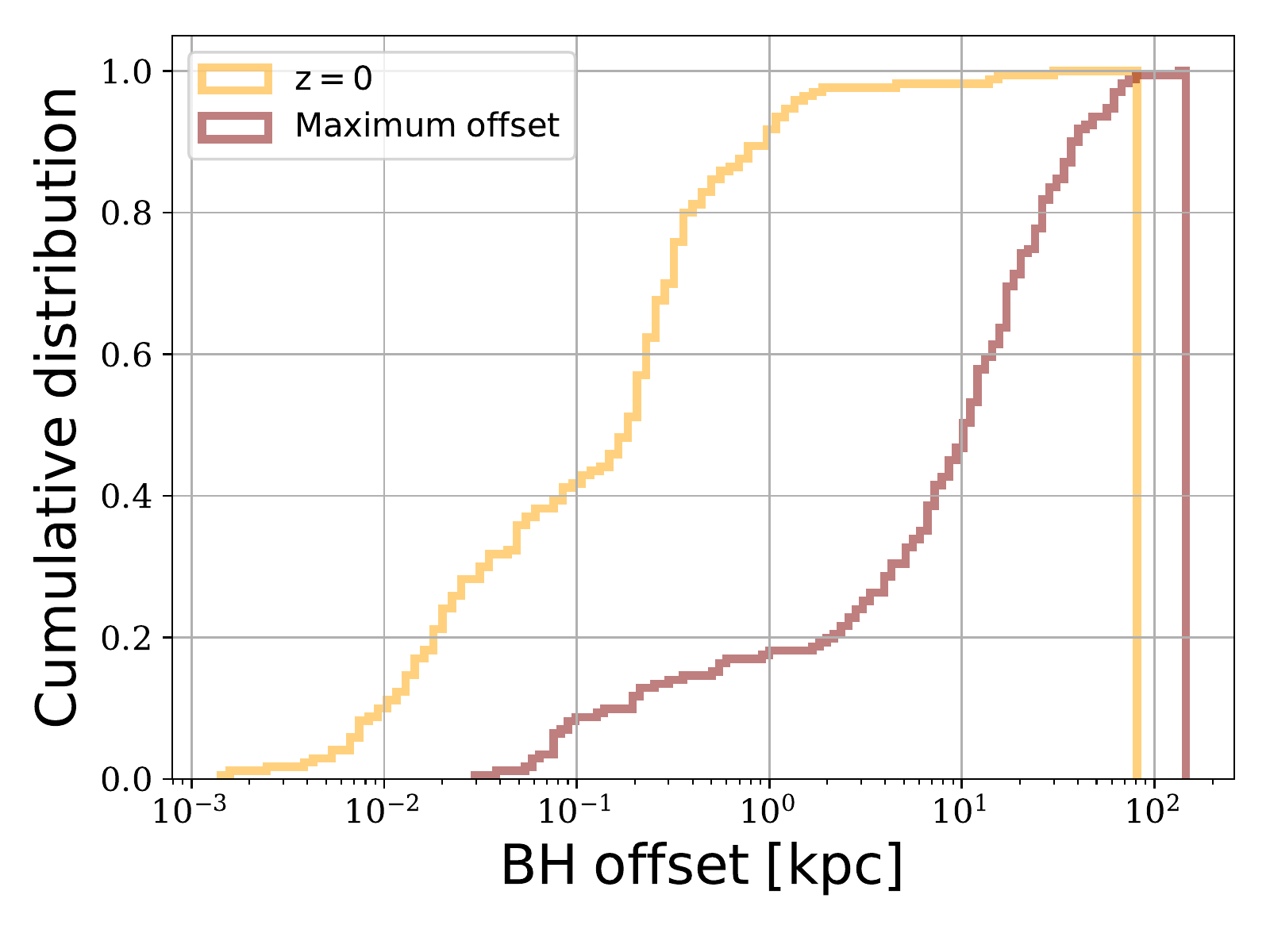}
\caption{\textit{Off-centered black holes in BCGs:} Cumulative histograms showing the BH offset at $z=0$ in yellow, and the maximum offset measured since $z=2$ in maroon. In almost half of our sample (170 BCGs), the SMBH is still off-centered ($r > 0.01$ kpc) at $z=0$. About 60$\%$ of these SMBHs are located at $r > 0.1$ kpc at present time, and more than 80$\%$ of these SMBHs were off-centered by at least 1 kpc during their dynamical history.}
\label{fig2}
\end{figure}

\begin{figure}
\centering
\includegraphics[width=\hsize]{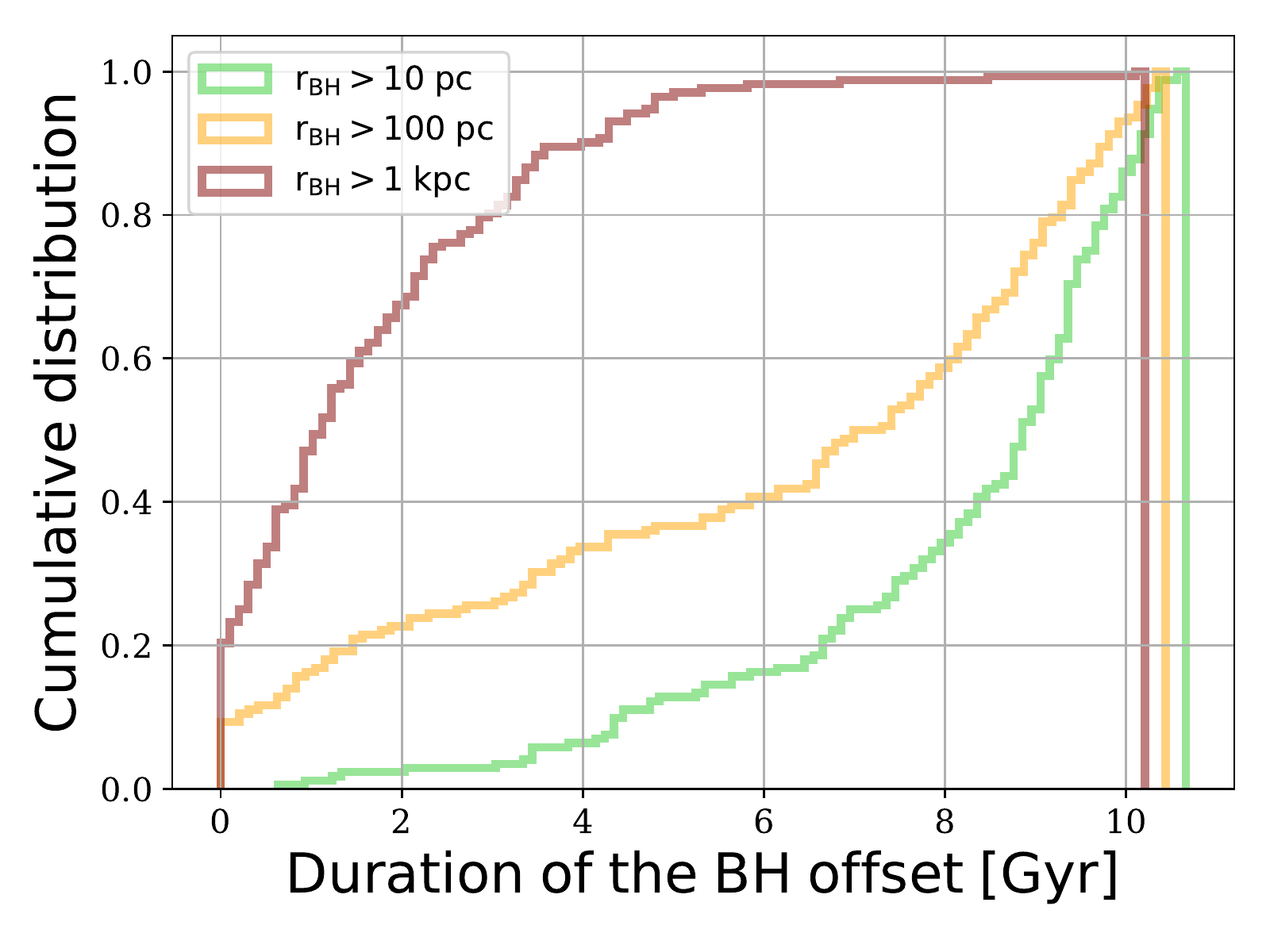}
\caption{\textit{Black hole offset time:} Cumulative histogram of the total duration of our 170 BHs with an offset $>$ 10, 100 and 1000 pc since $z=2$. A huge percentage 85$\%$ (60$\%$) of the BCGs exhibit a SMBH off-centered with $r>$ 10 pc ($r>$ 100 pc) for at least 6 Gyr. Besides, these BHs spend less than 3 Gyr above 1 kpc.}
\label{fig3}
\end{figure}

\begin{figure}
\centering
\includegraphics[width=\hsize]{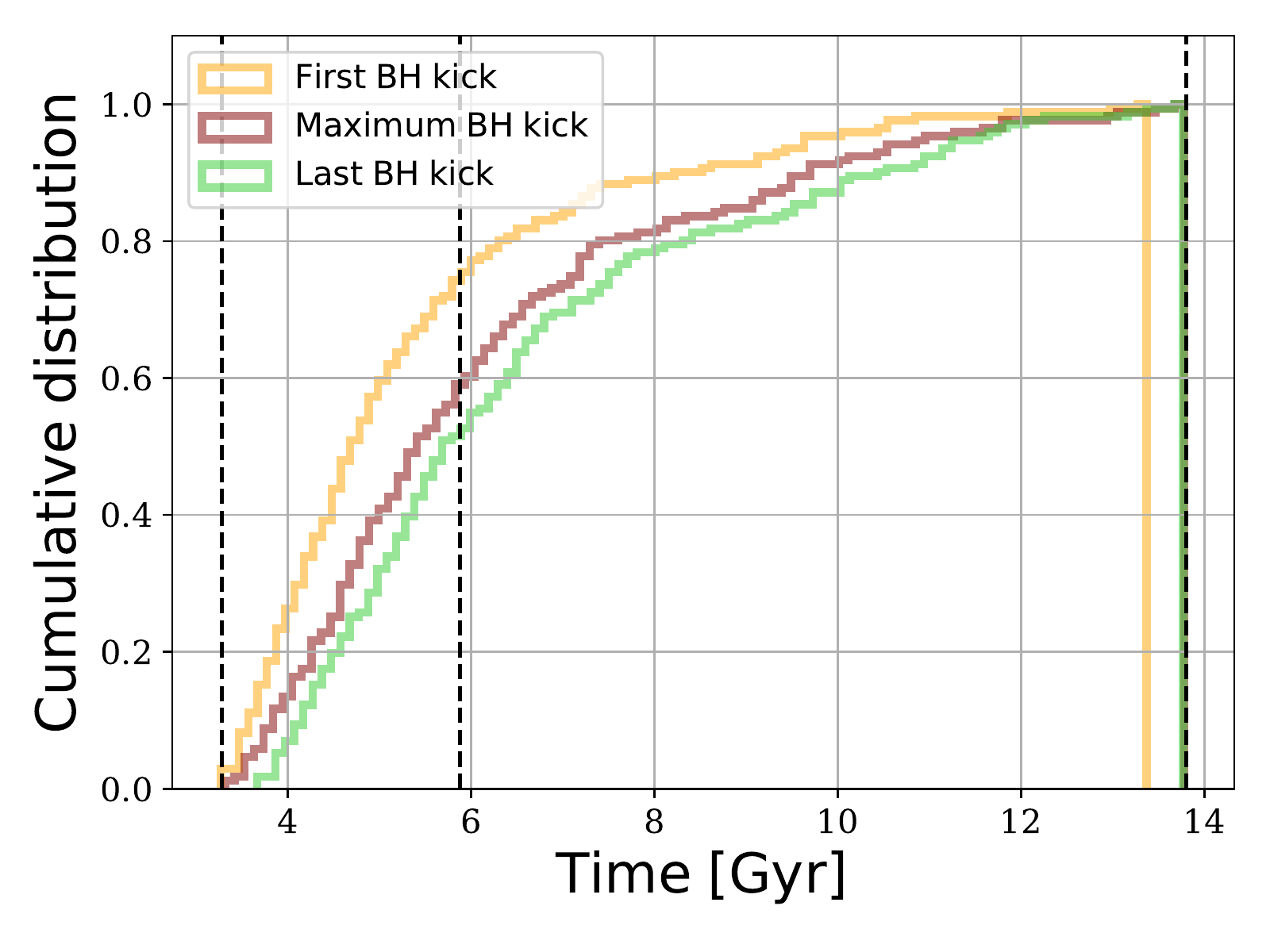}
\caption{\textit{Offset epochs:} Cumulative histograms showing the times when the first (yellow), the maximum (maroon) and the last (green) kick happened. The vertical black dashed lines show $z=2$, 1 and 0. 70$\%$ of our 170 BHs experienced their first kick before $z=1$. The merger which kicked the central SMBH to the maximum distance does not take place at a particular time since $z=2$, but still happens much earlier than the last merger. 65$\%$ of the clusters have undergone their last mergers after $z=1$. That is the reason why it is very likely that SMBHs in BCGs are still off-centered at $z=0$.}
\label{fig4}
\end{figure}

\begin{figure}
\centering
\includegraphics[width=\hsize]{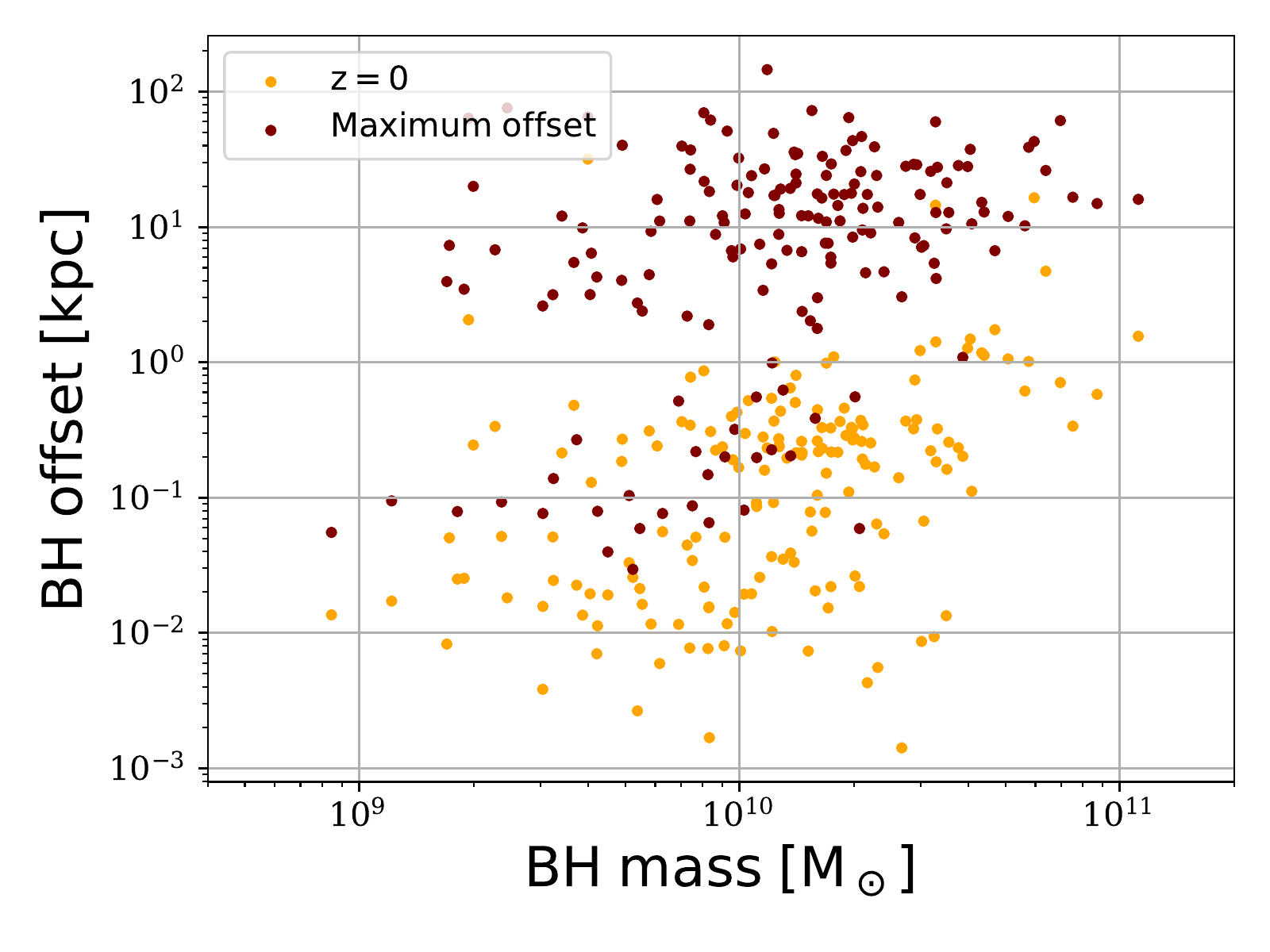}
\caption{\textit{Dynamical friction on black holes:} BH offset as a function of its mass at $z = 0$ (yellow dots) and at the time of the maximum offset (maroon dots). More massive BH exhibits larger offsets than less massive ones because dynamical friction acting on BHs becomes significantly weaker as the total galaxy mass grows.}
\label{fig5}
\end{figure}

Here we present our results by considering our full sample of 370 BCGs and thus 370 orbiting BHs, tracked with orbital integrations.

As depicted by Figure~\ref{fig1a}, not all mergers undergone by the BCGs perturb the central BH. We derive that BHs are mainly kicked by satellites which have stellar masses $M^{\rm sat}_{*}>M^{\rm BCG}_{*}/100$. We confirm that massive satellites on a radial orbit are rare (only 229 mergers out of all 6628 mergers accounted for all 370 clusters between $z=2$ and $z=0$ meet Equations \eqref{eq1} and \eqref{eq2}),
but their rareness does not invalidate their efficiency in displacing the BH of BCGs. Almost half of BHs ($\sim$170) of our sample were off-centered during their orbital history. Some of the more massive satellites also appear to not have decentered the BH, as indicated by the orange island of points above the line. Out of all mergers with a stellar mass ratio $M^{\rm sat}_{*}>M^{\rm BCG}_{*}/100$, 220 out of 3500 satellites kicked away the BHs from their central positions. Out of the 3280 satellites that did not manage to move the BH, 95\% are massive satellites which did not satisfy Equation \eqref{eq1}, i.e. their orbits are not radial. The remaining 5\% are massive satellites that do have a radial orbit, but the BH was still off-centered at the moment of the merger and was thus not affected. 
Thus, we predict two different populations of BHs: those that have always resided at the centre of their BCGs and those that have been off-centered. Hence, we expect that these BHs exhibit differences in accretion, growth and feedback. 

For all the 170 BHs heated by satellite galaxies, we analyse their orbit between $z=2$ and $z=0$. Figure~\ref{fig2} shows the cumulative distribution of the SMBH offset at $z=0$ and the maximum offset measured since $z=2$. In almost all these BCGs, the SMBH is still off-centered at $z=0$. About $60\%$ of these SMBHs are located at $r>0.1$ kpc at present time, and almost all SMBHs were off-centered by at least 1 kpc during their dynamical history. The maximum offset reached by these BHs is about 200 kpc (see Figure~\ref{fig2}). We also demonstrate that even if BHs have been significantly displaced at a specific redshift, BHs have sufficient time to decrease their offset. We stress that below 10 pc, the determination of SMBH orbits is not accurate as the dynamical friction becomes inefficient and three-body interactions with surrounding stars or other orbiting BHs can cause the BH orbit to further decay \citep{2015MNRAS.454L..66S,2017ApJ...840...31D,2018MNRAS.473.3410R}.

We now evaluate the time spent by the SMBH at distances longer than 10, 100 and 1000 pc from the BCG center. Figure~\ref{fig3} illustrates the cumulative histogram of the duration of the SMBH offset. We state that 85$\%$ (60$\%$) of the BCGs exhibit a SMBH with an offset of 10 pc (100 pc) for more than 6 Gyr since $z=2$. This means that SMBHs in BCGs spent more than half of their lifetime off-centered. We also notice that these BHs spend less than 3 Gyr above 1 kpc.

Besides, we show in Figure~\ref{fig4} the cumulative distribution of the specific times at which the first merger, the merger which has led to the maximum offset, and the last merger happened. We establish that 70$\%$ of our 170 BHs experienced their first kick before $z=1$. Hence, SMBHs do not inhabit the center of the BCG potential already at early times of the Universe. Most  BHs have undergone their most important merger in terms of dynamical heating, i.e. the one that displaces the most our BHs, after $z=1$, but much earlier than their last mergers. We then establish that the energy given by subsequent mergers is one of the key ingredient that permits BHs to reach these far distances of the order of a tens of kpc. More precisely, as a result of the first merger, BHs gain energy from the satellite by changing its velocity according to Equation~\eqref{eq3}. Before the second kick, BHs have eccentric orbits and thus are more energetic. This dynamical behaviour allows them to reach velocities which are higher than the escape velocity computed at their position, until large radius from the galaxy centre. Finally, we demonstrate that 65$\%$ of the clusters have undergone their last mergers after $z=1$. This phenomenon naturally explains why it is very likely that SMBHs in BCGs are still off-centered at $z=0$. Indeed, these BHs do not have a sufficient time to loose their angular momentum via dynamical friction and then reach the central region of the galaxy, where they are expected to reside. 

Finally, Figure~\ref{fig5} displays the BH offset as a function of the BH mass, both at $z$ = 0 and at the time the SMBH suffered its most significant kick. We recall that M$_{\rm BH}$ was fixed at the mass derived from the BCG mass at the first $T_{d}$. We report a correlation between BH mass and its offset. Indeed, more massive BH exhibits larger offsets than less massive ones at $z$ = 0. As a result, it is more likely to find off-centered SMBHs in the most massive BCGs. In the hierarchical structure formation model, the density at the centre decreases as the total galaxy mass grows. In fact, dynamical friction acting on BHs becomes significantly weaker and then BHs take more time to sink towards the centre of their BCGs, as demonstrated in Figure~\ref{fig5}.

\section{Discussion}
\label{section:discussion}

BH offsets of a few dozen parsecs were shown to have huge consequences on the physics that reign at the center of galaxies \citep{10.1093/mnras/sty2103,2022MNRAS.516..167B}. Even at such small distances, \citet{boldrini2020} have pointed out that accretion by the SMBH in dwarf galaxies would become inefficient, due to the lower density of gas at higher radii from the center. \citet{10.1093/mnras/sty2103} indicated that this might also be the case for more massive galaxies, such as BCGs. 

Our results clearly establish that SMBH offsets are common in BCGs as they have undergone frequent dynamical perturbations. Indeed, about a third of our SMBH sample are off-centred at the present time, and half of SMBHs were off-centered by at least 1 kpc during their dynamical history. Similarly to what was shown for dwarf galaxies \citep{bellovary2021_10.1093/mnras/stab1665}, we demonstrate that the reason for off-center locations is also mainly due to galaxy-galaxy mergers and more precisely mergers with satellites on radial orbits.
The orbital offset in BCGs can sustain up to at least 6 Gyr between $z=2$ and $z=0$ in half of our BCGs. Our result reinforces the prediction of a population of off-centered BHs, not only in dwarf galaxies \citep{1994MNRAS.271..317G,2014ApJ...780..187R,2011MNRAS.414.1127M,2004MNRAS.354..427I,2003ApJ...582..559V,2002ApJ...571...30S,bellovary2019_10.1093/mnras/sty2842,boldrini2020}, but also in very massive galaxies such as BCGs.

The fact that half of our SMBHs spend almost all their dynamical history offset from the BCG center should have major consequences on the BH and galaxy formation and evolution as there is an interplay between the feeding/feedback mechanisms of the SMBH and the host galaxy \citep{1998A&A...331L...1S}. In particular, it was pointed out that BHs must intersect with a highly accretable clump at some time in order to accrete gas efficiently \citep{10.1093/mnras/sty2103}. Hence, as off-centered BHs seems to be very common in BCGs, we expect that they accrete gas very inefficiently as gas clumps are centrally located. As a result, off-centered BHs should experience essentially no growth at all. It was already demonstrated that in cosmological simulations, SMBH offset has a dramatic effect on BH growth \citep{2022MNRAS.516..167B}. It is established that SMBHs gain mass via gas accretion \citep{1982MNRAS.200..115S} and via BH mergers. Even if BH mergers are strongly subdominant to gas accretion for SMBHs \citep{2022MNRAS.513..670N,2021RAA....21..212Z}, we pointed out that both mechanisms are effectively halted by mergers of satellites on radial orbits in BCGs. 

According to our results, we expect that growth and feedback for a non-negligible SMBH population in BCGs was quenched between $z=2$ and $z=1$ until $z=0$. For this reason, we predict the presence of a population of SMBHs with very different masses. It will result in a larger scatter in mass compared to what cosmological simulations predict. However, observations of BHs show the presence of a scatter around the M$_{\rm BH}$ - M$_{\rm BCG}$ relation \citep{2019ApJ...884..169G,2018ApJ...852..131B,2016ApJ...817...21S,2017MNRAS.464.4360M,2013ApJ...764..184M}. Such scatter is evidence of the complex co-evolution of SMBHs with their host galaxies in clusters \citep{2021NatRP...3..732V}. In particular, we establish that the growth of these SMBHs seems to be governed by one property of their host clusters, i.e. the number of satellites on radial orbits. 

The absence of gas accretion by the SMBHs will also obviously influence their feedback \citep{2014ARA&A..52..589H}. If the SMBHs are not accreting, BH feedback also becomes inefficient. Nevertheless, it is well established that the BH feedback can alter the DM distribution of galaxies. In fact, the repeating episodes of gas ejection and recycling can flatten the cuspy DM profile until the formation of a constant density in the central region of galaxies, called the DM core. This effect on the DM profile has been studied in massive galaxies in the Horizon-AGN simulation \citep{2017MNRAS.472.2153P,2019MNRAS.483.4615P} and in the NIHAO-AGN simulations \citep{2020MNRAS.495L..46M}.They showed that BH feedback can very slightly flatten the DM profile of BCGs. Our result, i.e the quenching of BH feedback due to the BH offset, reinforces the prediction concerning the inefficiency of this mechanism to create DM cores in massive galaxies such as BCGs.

This work highlights a major problem encountered in simulations which pin the SMBH to the center of the potential of the cosmological box \citep[][]{bahe2021_https://doi.org/10.48550/arxiv.2109.01489}. Doing so has many advantages: keeping the BH at the center renders sub-grid AGN models effective, and this ensures that the initial merger-driven growth of BHs can proceed rapidly as this early phase is highly uncertain. This also avoids the question of dynamical friction, which is still treated poorly within simulations \citep[][]{morton2021arXiv210315848M}. However this results in higher accretion rates, overestimated BH growth and higher counts of BH-BH mergers \citep[see][and references therein]{bahe2021_https://doi.org/10.48550/arxiv.2109.01489,Barausse2020ApJ...904...16B}. We confirm that BH repositioning in simulations \citep[see Section 2.2 of][which details the repositioning of the BH in the TNG simulation,]{Weinberger2017MNRAS.465.3291W} is nonphysical and show the need of new repositioning and dynamical friction models in order to correctly model the dynamics in galaxy clusters. There have now been many efforts on testing and incorporating dynamical friction from gas and stars in cosmological simulations \citep[e.g][]{tremmel2015MNRAS.451.1868T, 2019MNRAS.486..101P,Chen2022MNRAS.510..531C}, which do produce off-center SMBHs.

\section{Conclusion}
\label{section:conclusion}

Making use of the work of \cite{2018MNRAS.481.1809B}, we study how satellites affect the dynamics in the central region of 370 different BCGs from the TNG300 simulation, between redshift $z$ = 2 and $z$ = 0. By using orbital integration methods, we show how galaxies which fall in the potential of the BCG can heat dynamically the center of BCGs, and kick away the central SMBH to distances from a few parsecs up to a hundred kiloparsecs. Our study demonstrates that half of SMBHs in BCGs, more particularly those in the most massive galaxies, were kicked away since $z$ = 2, but only a third are still off-centered by more than 200 pc today. These BHs spent most of their lives outside the central region of their host, which has consequences on the physics that reign at the center of clusters. 

As gas is mostly condensed in the center of galaxies, BHs, if offcentered, do not encounter as many gas clumps. As a result, BHs can not accrete gas efficiently. Consequently, the growth of SMBH may have been overestimated. Indeed, the lack of efficient accretion means that they do not gain much mass in their lifetime. As the central SMBH can not accrete matter efficiently, AGN feeback also becomes inefficient. 

Following this study, analyzing how mergers can affect the distribution of DM in the central regions of BCGs might also be an interesting subject of research. Indeed, the multiple passage of satellites in the central region of BCGs may be responsible for ejecting DM out of the center of the galaxy, resulting in the flattening of the central DM density.

The presence of such a population of offcentered BHs in BCGs brings up the question of their detections and the identification of BCGs which may host a SMBH which has been kicked away in the recent past. Traces in the morphology of the BCG such as stellar tails or streams may hint at the passage of a massive satellite and thus at a recent merger. Observations of the central regions of these galaxies with integral field units such as the Multi Unit Spectroscopic Explorer (MUSE) may enable us to pinpoint the location of a potential offcentered BH by looking for the presence of a broadline region, characteristic of Seyfert galaxies or associated with the presence of an AGN \citep[][]{GASKELL2009140}. This is still a challenging project as \citet{Elitzur201410.1093/mnras/stt2445} show that the broadline emission decreases as the accretion rate of the BH decreases as well. If offcentered, we would thus not expect an intense emission.

In this paper, we only considered the central BH of the BCG. However, satellites may harbor a SMBH as well in their centers. We could thus expect a population of wandering BHs of different masses in the BCG. The characterization of this population of BHs and the knowledge of their orbits in the potential of the BCGs may give us a better idea of the rate of BH-BH mergers observed in the future.

\section*{Acknowledgements}

We thank Josh Borrow and Mark Vogelsberger for providing the Illustris data, and for useful comments. We also thank Florence Durret, Warren Massonneau and Yohan Dubois for fruitful discussions concerning this project. Finally, we thank the referee for his constructive comments and suggestions. This work was supported by the EXPLORAGRAM Inria AeX grant.

\section*{Data Availability}

The data that support the plots within this paper and other findings of this study are available from the corresponding author upon reasonable request.



\bibliographystyle{mnras}
\bibliography{example} 

\appendix

\section{Dynamical friction model}
\label{appDF}

\begin{figure}
\centering
\includegraphics[width=\hsize]{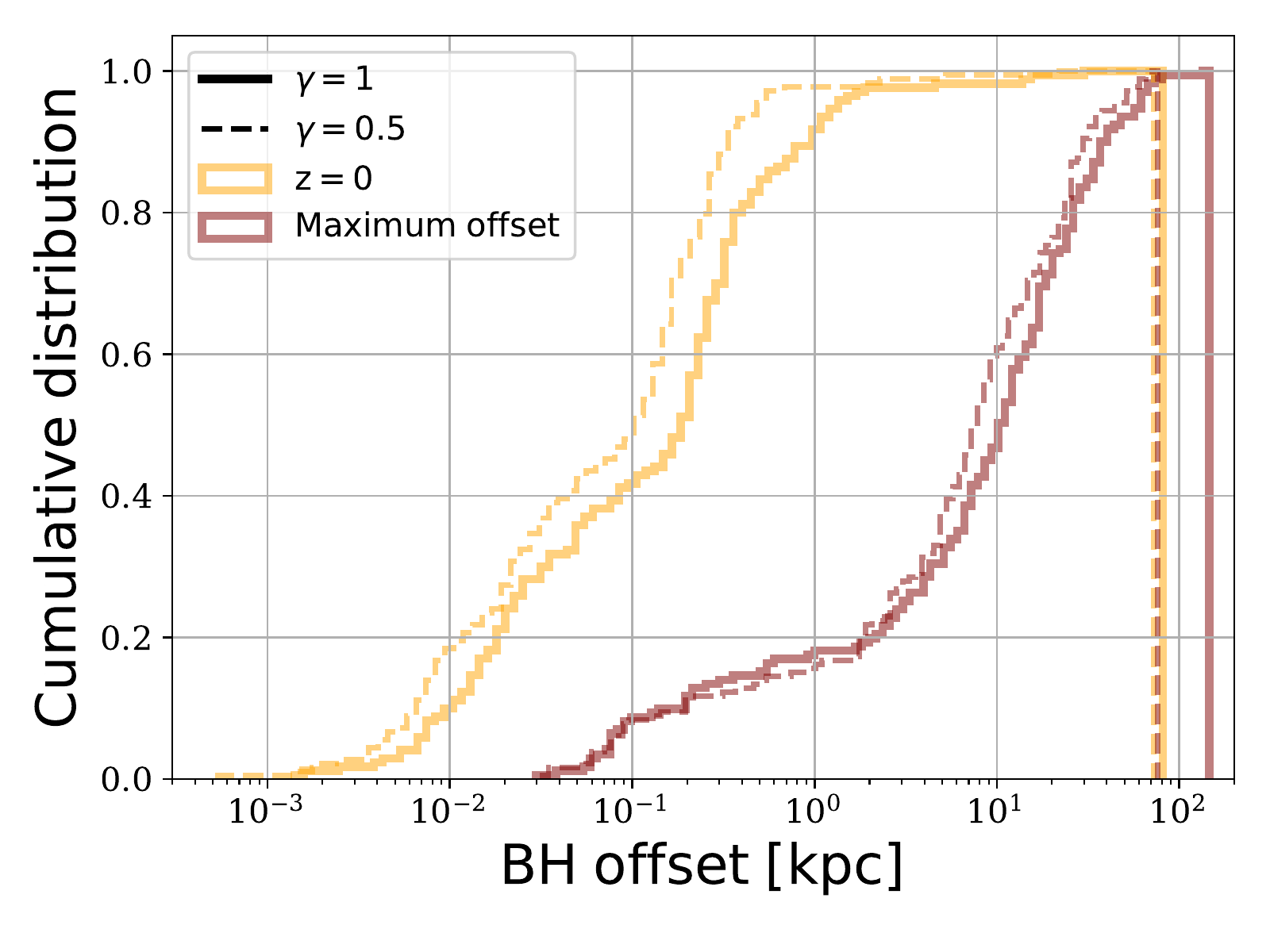}
\caption{\textit{Impact of the $\gamma$ parameter:} Cumulative histograms showing the BH offset at $z=0$ in yellow, and the maximum offset measured since $z=2$ in maroon. About 55$\%$ (60$\%$) of these SMBHs are located at $r > 0.1$ kpc at present time, and more than 80$\%$ (80$\%$) of these SMBHs were off-centered by at least 1 kpc during their dynamical history for $\gamma=0.5$ ($\gamma=1$).}
\label{fig1A}
\end{figure}

In this appendix, we discuss the sensitivity of the dynamical friction model chosen in our orbital integrations to track the orbit of central SMBHs in BCG potentials. The dynamical friction model is described in section~\ref{dfsection}. In addition, we evaluate here the impact of the $\gamma$ parameter, defined as the absolute value of the logarithmic slope of the density profile of BCGs. Indeed, this parameter controls the amplitude of the dynamical friction force during orbital integrations. In our study, $\gamma$ was set to 1. According to Figure~\ref{fig1A}, we observe a very slight deviation in both cumulative histograms by decreasing the $\gamma$ parameter. This figure demonstrates that modifying this dynamical friction parameter does not significantly affect our results.

\section{Assumption of fixed satellite mass}
\label{appML}

\begin{figure}
\centering
\includegraphics[width=\hsize]{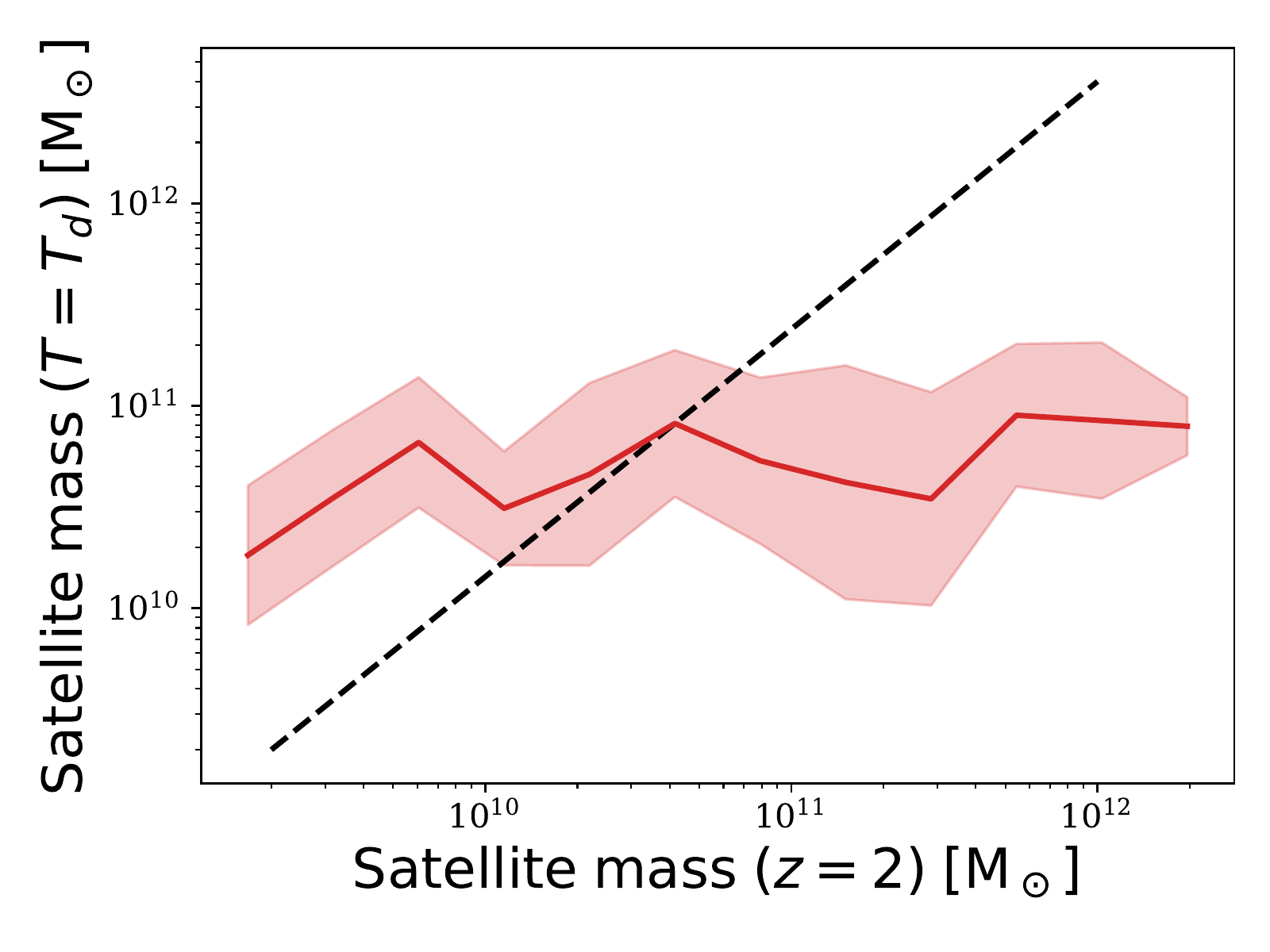}
\caption{\textit{Absence of mass loss during integrations:} Mean satellite mass at $T_{d}$ as function the satellite mass at $z=2$ for all the first mergers in BCGs. We display the mean over the satellite mass range at $z=2$, for better visualisation. Error bars represent 1-$\sigma$ Poisson uncertainties.}
\label{fig2A}
\end{figure}

In order to test our assumption of fixed satellite mass during the orbital integrations, we measure the mass loss of satellite galaxies between $z=2$ and the time of the first BH kick ($T_{d}$) in Illustris-TNG simulation. Figure~\ref{fig2A} depicts that below $10^{10}$ M$_{\sun}$ (above $10^{11}$ M$_{\sun}$) for the satellite mass at $z=2$, we over(under)estimated the satellite mass during our integrations by a factor of 10. Considering higher(lower) masses for satellites will affect their orbit by accelerating(delaying) their infall towards the BCG center. Taking into account the mass loss in orbital integration remains a challenging problem as it is computationally costly and also difficult to estimate due to the lack of resolution of the Illustris-TNG300 snapshot.

\bsp	
\label{lastpage}
\end{document}